\documentclass[aps,twocolumn,amssymb,showpacs,amsmath,amssymb,nfootinbib,preprintnumbers]{revtex4}

\usepackage{graphicx,bm,color}

\def\bra#1{\langle #1|}
\def\ket#1{|#1\rangle}

\begin{document}
\vspace{5cm}

\title{Particle creation in Bose--Einstein condensates: Theoretical formulation based on conserving gapless mean field theory}


\author{Yasunari Kurita$^{1}$, \ Michikazu Kobayashi$^{2}$, \ Hideki Ishihara$^3$, \ Makoto Tsubota$^3$}

\affiliation{$^1$
Kanagawa Institute of Technology, Atsugi, Kanagawa 243-0292, Japan}

\affiliation{$^2$
Department of Pure and Applied Sciences, The University of Tokyo, 3-8-1 Komaba, Tokyo 153-8902, Japan}

 \affiliation{$^3$
 Department of Physics, Osaka City University, Osaka 558-8585, Japan}

\vspace{2cm}

\begin{abstract}
We formulate particle creation phenomena in Bose--Einstein condensates
in terms of conserving gapless mean field theory for weakly interacting Bose gases. 
The particle creation spectrum is calculated by rediagonalizing the Bogoliubov--de Gennes (BdG) Hamiltonian in mean field theory.
The conservation implies that quasiparticle creation is accompanied by quantum backreaction to the condensates.
Particle creation in this mean field theory is found to be equivalent to that in quantum field theory (QFT) in curved spacetime.
An expression is obtained for an effective metric affected by quantum backreaction.
The formula for the particle creation spectrum obtained in terms of QFT in curved spacetime is shown to be the same as that given by rediagonalizing the BdG Hamiltonian.
\end{abstract}

\preprint{OCU-PHYS-342}
\preprint{AP-GR-86}


\maketitle
\section{Introduction}

Particle creation is one of the most important quantum phenomena predicted by quantum field theory (QFT) in curved spacetime;
in particular, it has important implications for cosmology and gravitational physics \cite{BD}.
When spacetime evolves dynamically, matter field quanta in 
spacetime are created even when no quanta initially exist.
Hawking radiation \cite{hawking} is the most well-known example
of particle creation. When a black hole is formed by gravitational collapse of a star, 
Hawking quanta are created and then emitted from the black hole as blackbody radiation.
Hawking radiation is commonly believed to cause a backreaction to the 
black-hole spacetime, resulting in black-hole evaporation.
However, a precise description of 
the backreaction of quantum particle creation to 
classical spacetime dynamics has not been developed yet \cite{Wald} 
(for recent progress see, for example, \cite{Maia:2007vy,Schutzhold:2008zzc}).

A scalar-field wave propagating in curved spacetime is analogous to a sound wave propagating in a fluid \cite{unruh81}.
In this analogy, background fluid flow corresponds to curved spacetime and provides the spacetime metric. 
The phase fluctuation of a sound wave can then be considered a field in curved spacetime.
This analogy enables us to use fluid systems to study many field theoretical phenomena in curved spacetime~\cite{novello2002,Barcelo:2005fc}.
It is also possible to use the analogy to derive some implications of 
quantum backreaction to spacetime dynamics by investigating condensed matter systems~\cite{Balbinot:2004da,Balbinot:2004dc,Ralf05,Balbinot:2006ua}.

One advantage of using such an analogy is the possibility of observing 
quantum effects experimentally.
It is important to consider quantum fluids to investigate quantum effects precisely.
Bose--Einstein condensates (BECs) in trapped cold 
atoms~\cite{Anderson-95,Ketterle95} are some of the best systems in this context \cite{Garay-PRL}.
Acoustic black holes has been recently reported for the first time in BEC~\cite{Lahav09}. 
One of the simplest methods for realizing a sonic horizon is just to expand a BEC \cite{Kurita:2007bv}.
The analogy of a BEC system is fascinating from a theoretical point of view because it provides 
a microscopic quantum description of quantum excitations. 
For example, a microscopic understanding of analog Hawking radiation has been obtained by investigating it in BECs in Bogoliubov theory 
\cite{Leonhardt,Recati,Macher:2009nz,Macher:2009tw}.

Particle creation and its quantum backreaction to condensates will occur in BEC 
systems and should be described by the theory of quantum fluctuations in condensed matter systems, 
independently of any analogy to curved spacetime.
Our aim is to formulate particle creation in condensed 
matter theory and to clarify the backreaction effect from 
quantum Bogoliubov quasiparticle creation to background condensates as 
well as its correction to particle creation itself.
This is important because backreaction effects are expected to be observed 
in experiments because of particle number conservation.

This aim will also be valuable for theoretical physics, if it succeeds 
in obtaining a method for dealing with the backreaction problem. 
For example, Sch\"utzhold et al.\cite{Ralf05} investigated quantum
backreaction in dilute
gas BECs, and pointed out that the effective-action technique does not
yield the
correct result in general.
This indication is important for gravitational physics, because
the prescription of quantum backreaction via the expectation value of the
pseudo-energy momentum tensor $T_{\mu\nu}$ to the Einstein equations,
which is usually considered, in the form
\begin{equation}
R_{\mu\nu}-\frac{1}{2}g_{\mu\nu} R=\langle {\hat T_{\mu\nu}} \rangle,
\end{equation}
would not be sufficient.
Therefore, the analogy provides a useful theoretical tool for 
considering the quantum aspect of the backreaction problem in QFT in 
curved spacetime.

The quantum backreaction to condensates has been discussed apart from the analogy context. 
Braaten and Nieto~\cite{Braaten-Nieto1997}, 
and Andersen and Braaten~\cite{Andersen-Braaten1999} discussed 
semiclassical corrections to a general time-independent condensate using 
Hartree--Fock--Bogoliubov (HFB) theory, 
which is a particle-number-conserving mean field theory.
Buljan {\it et al}.~\cite{Buljan05} considered the backreaction of 
thermal quantum fluctuations to condensate dynamics using HFB theory.
They performed a numerical simulation as an example of a backreaction in a finite-temperature BEC system.

We consider backreaction effects of particle creation to condensates 
which are not included in these works~\cite{Braaten-Nieto1997,Andersen-Braaten1999,Buljan05}. 
Particle creation is expected to occur in systems with dynamical 
background field (condensates or curved spacetime, and so on). 
Furthermore, particle creation occurs because of a change in definition of particles~\cite{BD}, 
and can not be evaluated only by solving time dependent 
Gross-Pitaevskii (GP) and Bogoliubov-de Gennes (BdG) equations, even when
these equations include backreaction terms as quasiparticle pair 
correlation. 
In the Bogoliubov theory, particle creation is formulated in terms of 
rediagonalization of the BdG Hamiltonian~\cite{Kurita09,Jannes10}, 
representing change in definition of quasiparticles.
Of course, in the Bogoliubov theory, the particle number of a condensate is conserved,
and the backreaction to the condensate cannot be taken into account.

HFB theory satisfies the conservation law within the total system.
Therefore, HFB theory is a candidate theory for treating the backreaction problem.
However, we do not consider this theory for the following reason:
The backreaction of particle creation has two aspects. 
One is the backreaction to the condensates, and 
the other is the backreaction to the particle creation spectrum itself, caused 
by the correction of the condensates. 
We explore a consistent formulation including these two aspects. 
For the purpose of evaluating the second aspect, 
a gapless theory is necessary. 
In a gapless theory, by rediagonalizing the BdG Hamiltonian, 
the condensate state can be considered in the calculation, and the mixing of the 
condensate state and the redefined quasiparticle states can be evaluated.
However, HFB theory is not gapless. 
Using HFB theory, we can not evaluate backreaction corrections to 
the particle creation spectrum.
Therefore, we do not consider HFB theory in the present article.
There is another reason why HFB theory is not available.
HFB theory has an energy gap in the excitation spectrum; 
 this contradicts the Hugenholtz--Pines theorem \cite{HP},
 which requires gapless excitation in the long-wavelength limit.

Recently, Kita \cite{Kita,Kita2,Kita3,Kita4} formulated an alternative mean field theory 
for BECs that both satisfies conservation laws 
and has gapless excitations.
It is thus a promising theory for describing BEC systems.
In this article, 
we consider the backreaction problem in terms of Kita theory.
By use of this theory, we formulate particle creation that includes 
backreaction effects in BECs at zero temperature (i.e., we do not consider 
thermal fluctuations). 
The particle creation spectrum corrected by backreaction effects is obtained.
We also formulate an analogy with curved spacetime in this theory.
By considering the analogy, we seek to obtain a model representing the 
quantum backreaction effect in the effective spacetime metric and to obtain 
some implications for QFT in curved spacetime.

The article is organized as follows.
In Sec. \ref{sec:review}, we review some basic equations and eigenstates of the BdG Hamiltonian in Kita theory, which are necessary for the subsequent discussion.
In Sec. \ref{sec:analogy}, we discuss particle creation based on QFT in curved spacetime.
For that purpose, we construct QFT in a spacetime analog based on Kita theory. 
In subsection \ref{sec:sub1}, the correspondence between the equation of motion for a quasiparticle field and a quantum field in curved spacetime is clarified.
In subsection \ref{sec:sub2}, a formula for the particle creation 
spectrum is obtained using QFT in a spacetime analog.
In Sec. \ref{sec:BEC-lang}, we formulate particle creation in BECs within BEC theory as rediagonalization of the BdG Hamiltonian.
The formula obtained for the particle creation spectrum is compared with that calculated in Sec. \ref{sec:analogy}.
Section \ref{sec:summary} gives the conclusions of the study.

\section{Review of basic equations: BdG eigenstates}
\label{sec:review}

We consider weakly interacting bosons trapped by an external confining potential $V_{\rm ext}({\bf x})$. 
The interaction is assumed to be a contact interaction with a strength denoted by $U_0$.
The field operator $\psi$ can be decomposed into the condensate wave function $\Psi$ and the quasiparticle field $\phi$
as 
\begin{eqnarray}
\psi =\Psi  + \phi. 
\end{eqnarray}
In time-dependent Kita theory, 
the equation for the condensate wave function $\Psi$
is given by
 \begin{eqnarray}
 i\hbar \frac{\partial \Psi}{\partial t} = 
 \left(
 \hat{L}+U_0 |\Psi|^2
 \right) \Psi +2U_0\langle\phi^{\dagger}\phi\rangle
 \Psi + U_0\langle\phi\phi\rangle\Psi^*, 
\label{eq:GP-Kita}
 \end{eqnarray}
where $\hat{L} = \hat{K} +V_{\rm ext}-\mu$.
$\hat{K} =-\frac{\hbar^2}{2m}\nabla^2$ is the kinetic energy operator and 
$\mu$ is the chemical potential. 
The extended Gross--Pitaevskii (GP) equation (Eq. (\ref{eq:GP-Kita})) includes 
quasiparticle pair correlations $\langle\phi^{\dagger}\phi\rangle$ 
and $\langle\phi\phi\rangle$.
These correlations are obtained from the renormalized Green's function~\cite{Kita4} 
and we assume that these quantities are finite. 
The condensate wave function $\Psi$ can be rewritten in terms of 
the condensate number density $n_0$ and the phase $S$ as:
\begin{eqnarray}
\Psi = \sqrt{n_0 }e^{iS}.
\end{eqnarray}
The total number of condensate particles is given by
\begin{eqnarray}
N_0 := \int |\Psi|^2 d^3x  = \int  n_0 d^3x .
\end{eqnarray}
In general, $N_0$ is time dependent in this theory.

The time-dependent BdG equation can be 
represented as
 \begin{eqnarray}
 i\hbar \partial_t  \mbox{\boldmath $\Phi$}
= {\bf H}  \mbox{\boldmath $\Phi$}
\label{eq:BdG-equation-Kita}
 \end{eqnarray}
where the BdG Hamiltonian is given by 
\begin{eqnarray}
{\bf H} =
\left(
\begin{array}{cc}
W+2U_0 \langle {\phi}^{\dagger}{\phi}\rangle
 & U_0 \Psi^2  - U_0 \langle {\phi}{\phi} \rangle 
\\
-U_0\Psi^{*2}  + U_0 \langle {\phi}^{\dagger}{\phi}^{\dagger} \rangle
& -W  -2U_0\langle {\phi}^{\dagger} {\phi}\rangle
\end{array}
\right),
\label{def:BdG-Hamiltonian}
\end{eqnarray}
where we have defined $W:=\hat{L}+2n_0U_0$.
$\mbox{\boldmath $\Phi$}$ is a two-component vector 
composed of $\phi$ and $\phi^{\dagger}$
\begin{eqnarray}
\mbox{\boldmath $\Phi$}(t,{\bf x}) = 
\left( 
\begin{array}{c}
\phi(t,{\bf x})\\
\phi^{\dagger}(t,{\bf x})
\end{array}
\right).
\end{eqnarray}
We introduce a similar vector for the condensate wave function,
\begin{eqnarray}
{\bf \Psi}(t,{\bf x}) =
\left( 
\begin{array}{c}
\Psi(t, {\bf x}) \\
\Psi^{*}(t,{\bf x})
\end{array}
\right).
\label{eq:condensate-wave-fnt-bogo}
\end{eqnarray}
It is found that $\hat{\tau}_3 {\bf \Psi}$ with the third Pauli matrix $\hat{\tau}_3$ is a solution of the BdG equation (Eq. (\ref{eq:BdG-equation-Kita})), which implies that the excitation spectrum is gapless.

Next, we review the eigenstates of the BdG Hamiltonian (\ref{def:BdG-Hamiltonian}) for some quasi-static situations. 
We denote the eigenvectors of the BdG Hamiltonian as $\mbox{\boldmath $\varphi$}_{m}$, which have two components.
They satisfy
\begin{eqnarray}
 {\bf H}  \mbox{\boldmath $\varphi$}_{m} =E_{m} 
 \mbox{\boldmath $\varphi$}_{m}, 
\end{eqnarray}
where $m$ is an integer.
As discussed in Ref. \cite{Kita2}, the orthonormal relation between eigenstates with positive eigenvalues can be assumed to be
\begin{eqnarray}
\langle \mbox{\boldmath $\varphi$}_{m}|\hat{\tau}_3|
\mbox{\boldmath $\varphi$}_{m'} \rangle = 
 \delta_{m m'}.
\label{eq:eq-ortho-bogo}
\end{eqnarray}
It can be shown that a positive eigenvalue $E_j$ of the BdG equation with the eigenvector
\begin{eqnarray}
{\bf u}_{j} := \left(
\begin{array}{c}
u_{j} \\ -v_{j}^* 
\end{array} \right), 
\label{eq:u-spinor-bogo}
\end{eqnarray}
is always accompanied by the negative eigenvalues $-E_j$ with the eigenvector
\begin{eqnarray}
{\bf v}_{j} := \left(
\begin{array}{c}
-v_{j} \\ u_{j}^* 
\end{array} \right). 
\label{eq:v-spinor-bogo}
\end{eqnarray}
Then, the normalization Eq.(\ref{eq:eq-ortho-bogo})
implies that 
\begin{eqnarray}
\langle u_{j} | u_{k} \rangle - \langle v_{k} | v_{j} \rangle =\delta_{jk}. 
\label{u-v-ortho-bogo}
\end{eqnarray}
Equivalently, the eigenstates with negative eigenvalues satisfy
\begin{eqnarray}
\langle {\bf v}_j |\hat{\tau}_3| {\bf v}_k \rangle= -\delta_{jk}.
\end{eqnarray}
In equilibrium, the BdG Hamiltonian has a zero eigenvalue whose eigenvector is proportional to $\hat{\tau}_3 | {\bf \Psi} \rangle$ with $| {\bf \Psi} \rangle$ 
given by Eq. (\ref{eq:condensate-wave-fnt-bogo}).
We normalize this eigenstate as follows:
\begin{eqnarray}
\mbox{\boldmath $\varphi$}_0 := 
\frac{1}{ 
\sqrt{2N_0}
}\hat{\tau}_3 {\bf \Psi} = \left(
\begin{array}{c}
\varphi_0  \\
-\varphi^{*}_0 
\end{array}
\right),
\label{eq:normaization-zero-spinor-bogo}
\end{eqnarray}
where we have defined the function $\varphi_0$ as
\begin{eqnarray}
\varphi_0  := \frac{1}{ 
\sqrt{2N_0}
}\Psi.
\label{eq:normaization-zero-fnct-bogo}
\end{eqnarray}
Noting that $\langle {\bf \Psi}| {\bf \Psi}\rangle = 2N_0$, 
we find $\langle \mbox{\boldmath $\varphi$}_0 |
 \mbox{\boldmath $\varphi$}_0 \rangle = 1$.
It is orthogonal to the states $|{\bf u}_j \rangle$ as 
\begin{eqnarray}
\langle {\bf u}_{j} | \hat{\tau}_3 |  \mbox{\boldmath $\varphi$}_0 \rangle =0,
\label{eq:ortho-zero-u-v-bogo}
\end{eqnarray}
As discussed in Ref. \cite{Kita2}, by noting Eq. (\ref{u-v-ortho-bogo}), 
we may assume 
\begin{eqnarray}
\langle u_{j} | \varphi_0 \rangle = \langle \varphi_0 | v_j 
 \rangle =0,
\label{eq:orthonormal-zero-mu-bogo}
\end{eqnarray}
giving
\begin{eqnarray}
\langle {\bf u}_{j} |  \mbox{\boldmath $\varphi$}_0 \rangle =0.
\label{eq:ortho-zero-u-v-bogo}
\end{eqnarray}
For BECs in equilibrium, these eigenstates represent Bogoliubov quasiparticles.

Using these eigenfunctions, 
the quasiparticle field operator can be expanded as
\begin{eqnarray}
\left(
\begin{array}{c}
\phi \\ \phi^{\dagger}
\end{array}
\right)
 &=& \sum_{k\ne0}
\left[
\hat{\gamma}_k \left(
\begin{array}{c}
u_{k} \\ -v_{k}^{*} 
\end{array} \right)  + 
\hat{\gamma}_k^{\dagger} \left(
\begin{array}{c}
-v_{k} \\ u_{k}^{*} 
\end{array} \right)
\right] \nonumber \\
&& + (\hat{\gamma}_0 -  \hat{\gamma}_0^{\dagger}) 
\left(
\begin{array}{c}
\varphi_{0}^{} \\ -\varphi_{0}^{*} 
\end{array} \right).
\label{eq:quaiparticle-expansion}
\end{eqnarray}
The operators $\hat{\gamma}_m$ and 
$\hat{\gamma}_n^{\dagger}$ ($m,n=0,1,2,\cdots$)
satisfy the boson commutation relation 
$[\hat{\gamma}_m, \hat{\gamma}_n^{\dagger}]=\delta_{mn}$.
The operators 
$\hat{\gamma}_k$ and $\hat{\gamma}_k^{\dagger}$ 
can be interpreted as annihilation and creation operators of 
quasiparticles, respectively.
$\hat{\gamma}_0$ can be interpreted as operators 
that annihilate a particle from the condensate, and 
$\hat{\gamma}_0^{\dagger}$ is the reverse operator. 
We have determined the minus sign before $\hat{\gamma}_0^{\dagger}$
in the last term of the expansion (Eq. (\ref{eq:quaiparticle-expansion})) 
in accordance with the discussion in Appendix \ref{sec:epsilon}.

\section{Analogy with QFT in curved spacetime}
\label{sec:analogy}

In this section, 
we discuss particle creation in BECs
using the analogy with QFT in curved spacetime.
For this purpose, we construct QFT in curved spacetime within 
the conserving gapless mean field theory \cite{Kita, Kita2,Kita3,Kita4}. 
First, we show that the equation of motion for quasiparticles 
can be viewed as an equation for a quantum field in curved spacetime.
The metric of the spacetime analog includes the backreaction effect. 
Next, we obtain the particle creation formula based on QFT in curved spacetime.

\subsection{field equation}
\label{sec:sub1}

The quasiparticle field can be U(1) gauge transformed as
 \begin{eqnarray}
 \tilde{\phi} := e^{-iS} \phi, \quad \tilde{\phi}^{\dagger}:=
 e^{iS}\phi^{\dagger}.
 \end{eqnarray}
The extended GP equation (\ref{eq:GP-Kita}) can be rewritten as the following two real equations:
  \begin{eqnarray}
 & \partial_t n_0 + \nabla \cdot (n_0 {\bf v}_0) - \frac{n_0 U_0}{i\hbar} 
   \left( \langle \tilde{\phi}\tilde{\phi}\rangle - \langle \tilde{\phi}^{\dagger}
 \tilde{\phi}^{\dagger}\rangle \right) =0, \nonumber \\
& \label{eq:extended-GP-real1} \\
 & \hbar \partial_t S + n_0^{-1/2} K n_0^{1/2}
 +V_{ext} -\mu +n_0 U_0 
 +\frac{1}{2}mv_0^2 \nonumber 
 \\
& +2U_0\langle\tilde{\phi}^{\dagger}\tilde{\phi}\rangle 
 + \frac{1}{2} U_0 \left( \langle\tilde{\phi}\tilde{\phi}\rangle +
 \langle \tilde{\phi}^{\dagger} \tilde{\phi}^{\dagger}\rangle \right) =0,
\label{eq:extended-GP-real2}
 \end{eqnarray}
where ${\bf v}_0 := \hbar \nabla  S/m$ is the velocity of the condensate.
The equations for $\tilde{\phi}$ and $\tilde{\phi}^{\dagger}$ are
 \begin{eqnarray}
 i\hbar \partial_t \tilde{\phi} 
 &=& 
 (\tilde{L}+2n_0 U_0+\hbar\partial_t S) \tilde{\phi} + n_0 U_0 \tilde{\phi}^{\dagger} 
 \nonumber \\
 &&
 +2U_0 \langle \tilde{\phi}^{\dagger}\tilde{\phi} \rangle \tilde{\phi}
 - U_0 \langle \tilde{\phi}\tilde{\phi} \rangle \tilde{\phi}^{\dagger},
 \label{eq:B-dG-1}\\
 -i\hbar \partial_t \tilde{\phi}^{\dagger}
 &=& 
 (\tilde{L}^*+2n_0 U_0 +\hbar  \partial_t S) \tilde{\phi}^{\dagger} 
 + n_0 U_0 \tilde{\phi}  \nonumber \\
  &&
 +2U_0 \langle \tilde{\phi}^{\dagger}\tilde{\phi} 
 \rangle \tilde{\phi}^{\dagger}
 - U_0 \langle \tilde{\phi}^{\dagger}\tilde{\phi}^{\dagger} \rangle \tilde{\phi},
 \label{eq:B-dG-2}
 \end{eqnarray}
where $\tilde{L}$ is defined as
 \begin{eqnarray}
 \tilde{L}:=\tilde{K}+V_{ext}-\mu, 
\end{eqnarray}
with
\begin{eqnarray}
\tilde{K}:=-\frac{\hbar^2}{2m}  \left(\nabla + i (\nabla S) \right)^2.
 \end{eqnarray}
Performing a short calculation, we obtain
\begin{eqnarray}
\tilde{K} &=& K -i\hbar {\bf v}_0\cdot 
\nabla 
-\frac{1}{2} i\hbar (\nabla \cdot {\bf v}_0)
+ \frac{1}{2}m v_0^2. 
\end{eqnarray}
Density fluctuations and current density fluctuations are
described by
\begin{eqnarray}
\hat{\rho}^{\prime} :=\sqrt{n_0}(\tilde{\phi}+\tilde{\phi}^{\dagger}), 
 \quad 
\hat{\bf j}' := \hat{\rho}^{\prime}{\bf v}_0+n_0 \hat{\bf v}^{\prime},
\end{eqnarray}
where we have introduced $\hat{\bf v}'=\nabla \hat{\Phi}'$
with 
\begin{eqnarray}
\hat{\Phi}':= \frac{\hbar}{2mi \sqrt{n_0}} 
(\tilde{\phi}-\tilde{\phi}^{\dagger}).
\end{eqnarray}
Taking the difference of Eqs. (\ref{eq:B-dG-1}) and (\ref{eq:B-dG-2}), 
we obtain the following conservation law
\begin{eqnarray}
\partial_t \hat{\rho}' + \nabla \cdot \hat{\bf j}^{\prime} =0. 
\label{eq:rho-1}
\end{eqnarray}
In the hydrodynamic approximation, in which the density gradient is 
smooth over the local healing length $\xi = {\hbar}/{\sqrt{2mU_0 n_0}}$, 
the sum of Eqs. (\ref{eq:B-dG-1}) and (\ref{eq:B-dG-2}) gives
\begin{eqnarray}
\hat{\rho}' \approx - \frac{m}{U_0X}
 \left(\partial_t +{\bf v}_0 
\cdot \nabla\right) \hat{\Phi}',
\label{eq:rho-approx}
\end{eqnarray}
where the quantity $X$ has been defined as 
\begin{eqnarray}
X:= 1 -  \frac{\langle\tilde{\phi}\tilde{\phi} \rangle 
+\langle \tilde{\phi}^{\dagger}\tilde{\phi}^{\dagger} \rangle  }
{2n_0}. 
\end{eqnarray}
Note that $X$ is real.
Then, Eq. (\ref{eq:rho-1}) can be rewritten as
\begin{eqnarray}
\partial_{\mu} \left[ \sqrt{-g}g^{\mu\nu} \partial_{\nu} \hat{\Phi}'\right]=0.
\end{eqnarray}
In this equation, we have defined the symmetric tensor $g_{\mu\nu}$ as
\begin{eqnarray}
g_{\mu\nu} := \frac{m c_s}{U_0 X} \left(
\begin{array}{cc}
-(c_s^2 -v_0^2) & -v_{0a} \\
-v_{0b} & \delta_{ab}
\end{array}
\right),
\label{eq:effectivemetric} 
\end{eqnarray}
where $a, b =1,2,3$.
The sound velocity can be described in terms of $X$ as
\begin{eqnarray}
c_s:= \sqrt{\frac{n_0 U_0 X}{m} }.
\end{eqnarray}
If we identify $g_{\mu\nu}$ with a spacetime metric through
$ds^2=g_{\mu\nu}dx^{\mu}dx^{\nu}$, the field $\hat{\Phi}^{\prime}$ 
satisfies the equation of motion for a minimally coupled scalar field
in curved spacetime with the metric given by Eq. (\ref{eq:effectivemetric}).
It can thus be considered as a field in curved spacetime.

The effective metric includes contributions from
quasiparticle pair correlations. 
If we drop the contributions and set $X=1$, then
the effective metric becomes the well-known metric 
without the backreaction (see, for example, \cite{Barcelo:2005fc}).
We find that the correction to the effective
metric appears in the form of $U_0X$. 
However, the backreaction effect is not 
included only in $X$: the quantities $n_0$ and $v_0$ 
are affected by the backreaction effect through the pair correlation 
terms in the extended GP equations, Eqs.
(\ref{eq:extended-GP-real1}) and (\ref{eq:extended-GP-real2}).

\subsection{Particle creation spectrum}
\label{sec:sub2}

We calculate the particle creation spectrum by using QFT in curved spacetime in the usual way, namely without considering a specific prescription for treating the backreaction to the background field.

First, we introduce U(1) gauged wave functions $\tilde{u}_j$ and $\tilde{v}_j$, which satisfy the U(1) gauged BdG equation
\begin{eqnarray}
i\hbar \partial_t\left( 
\begin{array}{c}
\tilde{u}_j \\ -\tilde{v}_j^*
\end{array} \right)
 =  \tilde{{\bf H}}
\left( \begin{array}{c}
\tilde{u}_j \\ -\tilde{v}_j^*
\end{array}  \right),
\label{eq:BdG-mode-u-v}
\end{eqnarray}
with the gauged Hamiltonian 
\begin{eqnarray}
\tilde{{\bf H}}:=\left(
\begin{array}{cc}
\tilde{W}+2U_0 \langle \tilde{\phi}^{\dagger}\tilde{\phi}\rangle
 & n_0 U_0 - U_0 \langle \tilde{\phi}\tilde{\phi} \rangle \\
-n_0U_0 + U_0 \langle \tilde{\phi}^{\dagger}\tilde{\phi}^{\dagger} \rangle
& -\tilde{W}^*-2U_0\langle \tilde{\phi}^{\dagger} \tilde{\phi}\rangle
\end{array}  \right),
\label{def:modified-BdG-Hamiltonian}
\end{eqnarray}
where $\tilde{W}:=\tilde{L}+2U_0 n_0+\hbar\partial_t S$. 
The normalized and gauged condensate wave function is defined by
\begin{eqnarray}
\tilde{\varphi}_0 := 
\frac{1}{\sqrt{2N_0}}\sqrt{n_0}.
\label{def:tilde-varphi-0}
\end{eqnarray}
Similarly, we define the gauged condensate wave function as
$\tilde{\Psi}:=e^{-iS}\Psi = \sqrt{n_0}$. 
The normalization factor $N_0$ in Eq. (\ref{def:tilde-varphi-0})
is the number of condensate particles at some time and it is a constant.
It is found that 
\begin{eqnarray}
\hat{\tau}_3 \left(\begin{array}{c}
\tilde{\varphi}_0 \\  \tilde{\varphi}_0^*
\end{array}\right) =  
\left(\begin{array}{c}
\tilde{\varphi}_0 \\ - \tilde{\varphi}_0^*
\end{array}\right) 
\end{eqnarray}
also satisfies the gauged BdG equation with the Hamiltonian 
Eq. (\ref{def:modified-BdG-Hamiltonian}).

With these fields, 
the U(1) gauged field $\tilde{\phi}=e^{-iS}\phi$ can be expanded as
\begin{eqnarray}
\tilde{\phi} = \sum_j \left( \hat{\gamma}_{j}\tilde{u}_j
-\hat{\gamma}_j^{\dagger} \tilde{v}_j \right)
+\hat{\gamma}_0 \tilde{\varphi}_0-\hat{\gamma}_0^{\dagger}\tilde{\varphi}_0.
\end{eqnarray}
The scalar field $\hat{\Phi}'$ can be expanded in terms of wave functions 
\begin{eqnarray}
\bar{f}_{j} = \frac{\hbar (\tilde{u}_{j}+\tilde{v}_{j}^*)}{2mi\sqrt{n_0}},
\label{def:bar-f-j}
\end{eqnarray}
for positive $j$ and 
\begin{eqnarray}
\bar{f}_0 &=& \frac{\hbar}{2mi\sqrt{n_0}} 
 (\tilde{\varphi}_0+\tilde{\varphi}_0^*) = \frac{\hbar}{mi\sqrt{2  N_0}},
\end{eqnarray}
as 
\begin{eqnarray}
\hat{\Phi}^{\prime} = \sum_{n=0}^{\infty} \left(
\hat{\gamma}_n \bar{f}_n + \hat{\gamma}_n^{\dagger}\bar{f}_n^*
\right). 
\end{eqnarray}
The zero mode $\bar{f}_0$ is a constant.

Under the hydrodynamic approximation, 
it can be shown that the mode functions $\{\bar{f}_n\}$ obey the equation of motion
in the spacetime analog 
\begin{eqnarray}
\partial_{\mu} \left[ \sqrt{-g}g^{\mu\nu} \partial_{\nu} \bar{f}_n \right]=0.
\label{eq:wave-equation-metric-form}
\end{eqnarray}
The details of this derivation are given in Appendix \ref{sec:equation-mode}.

Next, we investigate the Klein-- Gordon (KG) inner product between the 
mode functions.
The KG product for functions $F$ and $G$ is defined as
\begin{eqnarray}
(F,G)_{KG} = -i \int_{\Sigma}
\left(
Fn^{\mu}\partial_{\mu} G^* - G^* n^{\mu} \partial_{\mu} F
\right)\sqrt{\mbox{det}\ h} d^3x, \nonumber \\
\end{eqnarray}
where $n=n^{\mu}\partial_{\mu}$
is a unit normal vector orthogonal to the spacelike surface $\Sigma$; 
that is, $n$ is a future-directed timelike vector.
Additionally, 
$\sqrt{{\rm det}\ h}d^3x$ is an invariant volume element of the surface 
$\Sigma$, 
and $h_{ab}$ is the induced metric on $\Sigma$. 
If we choose a $t$-constant surface as $\Sigma$, the induced metric is
\begin{eqnarray}
ds^2_{\Sigma} = h_{ab}dx^adx^b = \frac{mc_s}{U_0\sqrt{X}}\delta_{ab} dx^adx^b.
\end{eqnarray}
Then, the determinant of the induced metric is
$\sqrt{{\rm det}\ h}=({mn_0}/{ U_0 X})^{3/4}$. 
The normal vector to the surface is
\begin{eqnarray}
n =  \left(\frac{m}{n_0^3U_0X}\right)^{1/4}
(\partial_t + {\bf v}_0 \cdot \nabla).
\end{eqnarray}
For the constant $\bar{f}_0$, it is trivial to show that 
\begin{eqnarray}
(\bar{f}_0, \bar{f}_0)_{KG} =0.
\label{eq:zero-zero-KG}
\end{eqnarray}
Furthermore, 
\begin{eqnarray}
(\bar{f}_j, \bar{f}_0)_{KG} 
 &=& \sqrt{\frac{{\hbar}}{{m}}} \int \sqrt{\frac{n_0}{2N_0}}
 (\tilde{u}_j-\tilde{v}_j^*) d^3 x \nonumber \\
&=& \sqrt{\frac{\hbar}{m}} \bigg(
\langle \varphi_0 | u_j \rangle - \langle v_j | \varphi_0 \rangle 
\bigg).
\label{eq:zero-j-KG}
\end{eqnarray}
When the system is quasi-static, Eq. (\ref{eq:orthonormal-zero-mu-bogo})
can be applied to the preceding equation, giving
\begin{eqnarray}
(\bar{f}_j, \bar{f}_0)_{KG} =0.
\label{eq:zero-j-KG-2}
\end{eqnarray}
Therefore, KG products including the zero-mode function $\bar{f}_0$ 
vanish in quasi-static situations.
The mode functions $\bar{f}_j$ with positive $j$ satisfy 
\begin{eqnarray}
n ^{\mu} \partial_{\mu} \bar{f}_{j} = 
- \left(\frac{m}{n_0^3U_0X}\right)^{1/4}
\frac{U_0X\sqrt{n_0}}{m } (\tilde{u}_{j}-\tilde{v}_{j}^*).
\label{eq:f-bar-u-v}
\end{eqnarray}
See Appendix \ref{sec:equation-mode} for the detailed calculation.
Using the orthonormal relation of Eq.(\ref{u-v-ortho-bogo}), it can be shown that
the functions are normal each other,
\begin{eqnarray}
(\bar{f}_{j}, \bar{f}_{k})_{KG} 
 &=& \frac{\hbar}{m} \delta_{jk}.
\end{eqnarray}
Similar calculations show that 
$(\bar{f}_j^*, \bar{f}_k^*)_{KG}=-\frac{\hbar}{m}\delta_{jk}$
and $(\bar{f}_j, \bar{f}_k^*)_{KG}=0$.
Therefore, mode functions with positive $j$ 
are normal each other.
At the same time, they are 
not orthonormal.
The mode functions should be orthonormal to identify the number of quanta in effective curved spacetime and the number of Bogoliubov quasiparticles.

For this purpose, we normalize the field $\hat{\Phi}^{\prime}$ as
\begin{eqnarray}
\hat{\varphi} =\frac{\sqrt{\hbar}}{2i\sqrt{mn_0}}(\tilde{\phi}-\tilde{\phi}^{\dagger})
=\sqrt{\frac{m}{\hbar}}\hat{\Phi}^{\prime}.
\label{eq:field-curved}
\end{eqnarray}
With the normalized mode functions 
\begin{eqnarray}
f_j:= \frac{\sqrt{\hbar}(\tilde{u}_j+\tilde{v}_j)}{2i\sqrt{mn_0}},
\end{eqnarray}
and 
\begin{eqnarray}
f_0 &=& \frac{\sqrt{\hbar}}{2i\sqrt{mn_0}} 
 (\tilde{\varphi}_0+\tilde{\varphi}_0^*) 
= \frac{\sqrt{\hbar}}{i\sqrt{m} N_0 },
\end{eqnarray}
the normalized field $\hat{\varphi}$ can be expanded as 
\begin{eqnarray}
\hat{\varphi} 
&=& \sum_{n=0}^{\infty} (\hat{\gamma}_n f_n + \hat{\gamma}^{\dagger}_n f_n^*). 
\label{eq:expansion-normalized-analogy-field}
\end{eqnarray}
The number operator of the field $\hat{\varphi}$ is given by
\begin{eqnarray}
\hat{N}_j = \hat{\gamma}_j^{\dagger}\hat{\gamma}_j,
\end{eqnarray}
which is the same as the number operator of Bogoliubov quasiparticles.
We have exactly related one Bogoliubov quasiparticle 
with one quantum in the spacetime analog.

As a simple example of particle creation, 
we consider a condensate that is quasi-static and has no quasiparticle
excitation at initial time $t=t_1$.
For $t_1 < t < t_2$, the condensate evolves dynamically and
becomes quasi-static again at $t = t_2$.
At the initial and final times $t=t_i$ (i=1,2), we expand
the field as
\begin{eqnarray}
\hat{\varphi} = \sum_{n=0}^{\infty} (\hat{\gamma}_n^{(i)}
 f_n^{(i)} + \hat{\gamma}^{(i)\dagger}_n f_n^{(i)*}),
\end{eqnarray}
where $\{f_n^{(i)}\}$ is a complete set at each time $t=t_i$.
Each $f_n^{(i)}$ consists of eigenfunctions of the BdG equations $u_j^{(i)}$,
$v_j^{(i)}$, and $\varphi_0^{(i)}$ at each time $t=t_i$.
With annihilation and creation operators $\hat{\gamma}_n^{(i)}$ and 
$\hat{\gamma}_n^{(i)\dagger}$,
we define one particle state of the quantum.

The calculation of the particle creation spectrum based on 
QFT in effective spacetime is exactly the same 
as that given in \cite{Kurita09}, except that 
the current expansion includes the zero mode $f_0^{(i)}$.
However, as in Eqs. (\ref{eq:zero-zero-KG}) and (\ref{eq:zero-j-KG-2}),
the zero mode does not contribute to the KG product, 
and does not alter the particle creation spectrum.
Therefore, the particle creation spectrum is the same as that given in Ref. \cite{Kurita09}.
Explicitly, the expectation value of the final number operator in 
the $j$th mode 
$\hat{N}_j^{(2)} := \hat{\gamma}_j^{(2)\dagger}\hat{\gamma}_j^{(2)}$ is 
\begin{align}
{}_{(1)} \bra{0} \hat{N}_j^{(2)} \ket{0}_{(1)} = \sum_k |B_{jk}|^2, 
\label{eq:formula-spectrum-analogy}
\end{align}
where the coefficient $B_{jk}$ is calculated as the KG product 
at the final time $t = t_2$,
\begin{align}
B_{jk} &= - (f_j^{(2)}, f_k^{(1)\ast})_{KG} \nonumber  \\
& = \int_{t=t_2} d^3x (u_j^{(2)} v_k^{(1)*}- v_j^{(2)*} u_k^{(1)}).
\label{eq:formula-B-analogy}
\end{align}
This expression agrees with the one obtained in Ref. \cite{Kurita09}, which 
did not consider any backreaction effect~
\footnote{To be precise, the notation for the negative energy eigenfunctions 
$v_j$ differs from that in Ref. \cite{Kurita09}.
To obtain identical expressions as those given in Ref. \cite{Kurita09}, replace $v_j \ with v_j^*$.}.
However, we should note that wave functions $u_j$, $v_j$, and $\varphi_0$ in Eq. (\ref{eq:formula-B-analogy}) are solutions of the equations that already include backreaction effects.
Therefore, although the expression for the particle creation is the same, 
it differs quantitatively from the result obtained in 
Ref. \cite{Kurita09} until the large condensate limit ($N_0 \to \infty$) is taken.

\section{Particle creation as BdG rediagonalization}
\label{sec:BEC-lang}

In this section, we consider particle creation within Kita theory 
as the rediagonalization of the BdG Hamiltonian, without referring to any effective spacetime.

Now, we consider the eigenspace of the BdG Hamiltonian.  
The unit operator that acts on the eigenspace is assumed to be 
expressed in terms of eigenstates as 
\begin{eqnarray}
 \hat{\bf 1} &=& \sum_{j}\bigg[
 | {\bf u}_j \rangle \langle {\bf u}_j | \hat{\tau}_3
 - 
 | {\bf v}_j \rangle \langle {\bf v}_j | \hat{\tau}_3
 \bigg]  + 
 | \mbox{\boldmath $\varphi$}_0 \rangle \langle \mbox{\boldmath 
 $\varphi$}_0 |. \nonumber \\
\label{eq:complete-set-equillibrium}
\end{eqnarray}
This complete set differs from that introduced in Ref. \cite{Kita2}, which includes a state that does not belong to the BdG eigenstates.
Since we are interested in solutions to the BdG equation, we assume the complete set given by Eq. (\ref{eq:complete-set-equillibrium}).

If the condensate and quasiparticle system are completely static, then no particle creation will occur. 
However, particles will be created in general when the condensate moves dynamically. 
As a simple example of particle creation, we consider the same situation for BECs as considered in Sec. \ref{sec:sub2},
namely the time evolution of condensate for $t_1<t<t_2$ accompanied 
by the initial and final quasi-static condensates.
Then, the complete sets that diagonalize the BdG Hamiltonian 
at $t=t_1$ and $t=t_2$ will be different in general, 
and annihilation and creation operators that define quasiparticle 
states will change.

The complete set that diagonalizes the BdG 
Hamiltonian at time $t=t_i (i=1,2)$ is assumed to be 
$\{|  {\bf u}_j^{(i)} \rangle, |  {\bf v}_j^{(i)} \rangle,
 | \mbox{\boldmath $\varphi$}^{(i)}_0 \rangle   \}$, 
satisfying
\begin{eqnarray}
 \hat{\bf 1} &=& \sum_{j}\bigg[
 | {\bf u}_j^{(i)} \rangle \langle {\bf u}_j^{(i)} | \hat{\tau}_3
 - 
 | {\bf v}_j^{(i)} \rangle \langle {\bf v}_j^{(i)} | \hat{\tau}_3
 \bigg] 
 + 
 | \mbox{\boldmath $\varphi$}^{(i)}_0 \rangle \langle 
\mbox{\boldmath  $\varphi$}^{(i)}_0 |. \nonumber \\
\end{eqnarray}
As for the condensate wave function, we have 
normalized the state $|\varphi_0^{(i)} \rangle$ 
at each time $t=t_i$ as
\begin{eqnarray}
{\varphi}_0^{(i)}(t_i,{\bf x}) := \frac{1}{\sqrt{2N_0^{(i)}}}
{\Psi}(t_i,{\bf x}),
\end{eqnarray}
with 
 \begin{eqnarray}
N_0^{(i)}: = 
\int n_0(t_i, {\bf x}) d^3x.
\end{eqnarray}
The particle number of the condensate will change through dynamic evolution.
$N_0^{(i)}$ is the particle number at each time and is a constant.

The quasiparticle field operator can be expanded as
\begin{eqnarray}
\left(
\begin{array}{c}
\phi \\ \phi^{\dagger}
\end{array}
\right)
 &=& \sum_{k\ne0}
\left[
\hat{\gamma}_k^{(i)} \left(
\begin{array}{c}
u_{k}^{(i)} \\ -v_{k}^{(i)*} 
\end{array} \right)  + 
\hat{\gamma}_k^{(i)\dagger} \left(
\begin{array}{c}
-v_{k}^{(i)} \\ u_{k}^{(i)*} 
\end{array} \right)
\right] \nonumber \\
&& + (\hat{\gamma}_0^{(i)} -  \hat{\gamma}_0^{(i)\dagger}) 
\left(
\begin{array}{c}
\varphi_{0}^{(i)} \\ -\varphi_{0}^{(i)*} 
\end{array} \right),
\label{eq:quaiparticle-expansion-at-each-time}
\end{eqnarray}
for $i=1,2$.
From the orthonormal relation at $t=t_2$, we obtain the following expansions:
\begin{eqnarray}
\hat{\gamma}_k^{(2)} &=& \sum_{j\ne0}
\left[
\hat{\gamma}_j^{(1)} \langle {\bf u}_k^{(2)} | \hat{\tau}_3 |{\bf 
u}_j^{(1)} \rangle
+ \hat{\gamma}_j^{(1)\dagger}\langle {\bf u}_k^{(2)}|\hat{\tau}_3
| {\bf v}_j^{(1)} \rangle \right] \nonumber \\
&& + (\hat{\gamma}_0^{(1)} - \hat{\gamma}_0^{(1)\dagger}  )
\langle {\bf u}_k^{(2)} |\hat{\tau}_3 | 
\mbox{\boldmath $\varphi$}^{(1)}_0  \rangle, 
\label{eq:gamma(1)-2-bogo}
\end{eqnarray}
and
\begin{eqnarray}
\hat{\gamma}_k^{(2)\dagger} &=& 
-\sum_{j\ne0}\left[
\hat{\gamma}_j^{(1)} \langle {\bf v}_k^{(2)} | \hat{\tau}_3 |{\bf u}_j^{(1)}\rangle
+ \hat{\gamma}_j^{(1)\dagger}\langle {\bf v}_k^{(2)}|\hat{\tau}_3 | 
{\bf v}_j^{(1)} \rangle \right] \nonumber \\
&&  - (\hat{\gamma}_0^{(1)} - \hat{\gamma}_0^{(1)\dagger}  )
\langle {\bf v}_k^{(2)} |\hat{\tau}_3 | 
\mbox{\boldmath $\varphi$}^{(1)}_0  \rangle. 
\label{eq:gamma(1)dagger-2-bogo}
\end{eqnarray}
It can easily be shown that the 
bra--kets in the preceding expressions satisfy the following relations:
\begin{eqnarray}
&&\langle {\bf u}_k^{(2)}|\hat{\tau}_3
| {\bf u}_j^{(1)} \rangle =
-\langle {\bf v}_k^{(2)} | \hat{\tau}_3 |{\bf v}_j^{(1)}\rangle^*, \\
&& \langle {\bf u}_k^{(2)}|\hat{\tau}_3
| {\bf v}_j^{(1)} \rangle =
-\langle {\bf v}_k^{(2)} | \hat{\tau}_3 |{\bf u}_j^{(1)}\rangle^*, \\
&& \langle {\bf u}_k^{(2)} |\hat{\tau}_3 | 
\mbox{\boldmath $\varphi$}^{(1)}_0  \rangle
= \langle {\bf v}_k^{(2)} |\hat{\tau}_3 | 
\mbox{\boldmath $\varphi$}^{(1)}_0  \rangle^*.
\end{eqnarray}

Since we have assumed that the initial condensate has no excitations, 
the quasiparticle state is the initial vacuum denoted by 
$|0\rangle_{(1)}$, satisfying $\hat{\gamma}_k^{(1)}|0\rangle_{(1)}=0$.
Then, the expectation value of the final number operator 
$\hat{N}_j^{(2)}:=\hat{\gamma}_j^{(2)\dagger}\hat{\gamma}_j^{(2)}$ 
is given by
\begin{eqnarray}
&& {}_{(1)}\langle 0 | \hat{N}_j^{(2)}
| 0 \rangle_{(1)} =
\sum_{k\ne0} | \langle {\bf u}_j^{(2)} |\hat{\tau}_3
| {\bf v}_k^{(1)} \rangle|^2 \nonumber \\
&& \qquad  - |\langle {\bf u}_j^{(2)} | \hat{\tau}_3 | 
\mbox{\boldmath $\varphi$}^{(1)}_0 \rangle|^2
{}_{(1)} \langle 0|(\hat{\gamma}_0^{(1)} 
- \hat{\gamma}_0^{(1)\dagger}
)^2 |0 \rangle_{(1)}. \nonumber \\
\label{eq:spectrum-formula-pre}
\end{eqnarray} 
The coefficient of the second term on the right-hand side vanishes for the following reason.
We have assumed that the system is quasi-static at $t=t_2$, then
the orthonormal relation between states implies that 
\begin{eqnarray}
\langle u_j^{(2)}|\varphi_0^{(2)} \rangle =
\langle v_j^{(2)} | \varphi_0^{(2)} \rangle =0.
\label{eq:orthonormal-t2-0-u-v}
\end{eqnarray}
Furthermore, the condensate wave functions $\varphi_0^{(2)}$ and $\varphi_0^{(1)}$
at time $t_2$ are related as
\begin{eqnarray}
\varphi_0^{(2)}(t_2,{\bf x}) = \frac{1}{\sqrt{2N_0^{(2)}}} \Psi(t_2,{\bf 
 x})
= \sqrt{\frac{N_0^{(1)}}{N_0^{(2)}} } \varphi_0^{(1)}(t_2,{\bf x}).
\label{eq:t1-t2}
\end{eqnarray}
Using Eqs. (\ref{eq:orthonormal-t2-0-u-v}) and (\ref{eq:t1-t2}),
it can be shown that
\begin{eqnarray}
\langle {\bf u}_j^{(2)} | \hat{\tau}_3 | 
\mbox{\boldmath $\varphi$}^{(1)}_0 \rangle =0.
\end{eqnarray}
Therefore, Eq. (\ref{eq:spectrum-formula-pre}) becomes
\begin{eqnarray}
&&
{}_{(1)} \langle 0 |  \hat{N}_j^{(2)}
| 0 \rangle_{(1)} 
=
\sum_{k\ne0} | \langle {\bf u}_j^{(2)} |\hat{\tau}_3
| {\bf v}_k^{(1)} \rangle|^2, 
 \label{eq:spectrum-formula}
\end{eqnarray} 
where
\begin{eqnarray}
\langle {\bf u}_j^{(2)} |\hat{\tau}_3
| {\bf v}_k^{(1)} \rangle 
 =
\int_{t=t_2} d^3x \left(
v_j^{(2)}u_k^{(1)*}-u_j^{(2)*} v_k^{(1)} 
\right) 
\label{eq:integral-t2-in-spectrum}
\end{eqnarray}
If the right-hand side of Eq. (\ref{eq:spectrum-formula}) 
is not zero, quasiparticles will appear at $t=t_2$.
This is the formula for the particle creation
spectrum in Kita theory.
The spectrum given by Eq. (\ref{eq:spectrum-formula}) 
includes the backreaction effect.
This expression agrees with Eq. (\ref{eq:formula-spectrum-analogy})
calculated by QFT in effective spacetime.

To evaluate the integral (\ref{eq:integral-t2-in-spectrum}) at $t=t_2$, 
one has to solve the time-dependent BdG equation to obtain the value of functions $u_k^{(1)}$ and $v_k^{(1)}$ at $t=t_2$.
Therefore, particle 
creation reflects the history of how the condensate evolves.

In the remainder of this section, we discuss the relation to the usual Bogoliubov theory,
which does not deal with the backreaction to condensates. 
To reduce the usual Bogoliubov theory, 
we need to neglect the correlation terms 
$\langle {\phi}^{\dagger}{\phi} \rangle$ and 
$\langle {\phi}{\phi} \rangle$ in 
the extended GP equation (Eq. (\ref{eq:GP-Kita})) and in the BdG equation 
(Eq. (\ref{eq:BdG-equation-Kita})).
The usual Bogoliubov theory is also gapless, and 
the condensate wave function is a solution of the BdG equation.
In the usual Bogoliubov theory, 
the particle creation spectrum can be derived in exactly the same manner 
as that given in this section. 
Of course, in the Bogoliubov theory, the particle number 
of the condensate is conserved, implying $N_0^{(1)}=N_0^{(2)}$.
This point is different, but it does not affect the result.
The expression for the particle creation spectrum 
given by Eq. (\ref{eq:spectrum-formula}) is also 
valid for the usual Bogoliubov theory.

The expression of the particle creation spectrum (Eq. (\ref{eq:spectrum-formula}))
agrees with that obtained in Ref. \cite{Kurita09}, which is 
calculated using the analogy with QFT in curved spacetime.
Therefore, the particle creation discussed in Ref. \cite{Kurita09}
can be thought of as a phenomenon explained by 
rediagonalization of the BdG Hamiltonian within the usual Bogoliubov 
theory.

\section{Conclusion and Discussion}
\label{sec:summary}

We have shown that particle creation in BEC
can be formulated as rediagonalization of the BdG Hamiltonian
in Kita theory.
The expression for the particle creation spectrum has been obtained, where
the quantum backreaction effects are taken into account.
Surprisingly, it is the same, apparently, as one obtained within Bogoliubov 
theory~\cite{Kurita09}, 
but the wave functions in Eq.(\ref{eq:integral-t2-in-spectrum}) obey the 
equation affected by the backreaction 
and are quantitatively different from those in Bogoliubov theory \cite{Kurita09}, in general.
The analogy with curved spacetime has been constructed within Kita theory.
By using QFT in effective spacetime, we have
obtained an expression for the particle creation spectrum, 
which gives the same spectrum as that obtained by rediagonalizing the BdG Hamiltonian.

This implies that the usual prescription for particle creation in QFT in 
curved spacetime might give the correct result.
In the analogy, the explicit backreaction effect appears only in the effective
spacetime metric.

The BdG Hamiltonian rediagonalization formulation provides a rigorous 
result for particle creation from a theoretical point of view, 
in the sense that the hydrodynamic approximation (which is assumed in constructing the analogy) is not required in this formulation.
Recently, Barcel\'{o}, Garay and Jannes \cite{Jannes10} formulated 
the analogy between Bogoliubov quasiparticles and quanta in curved spacetime
beyond the hydrodynamic approximation by introducing a generalized Klein-Gordon inner product. 
Using this inner product, the calculation based on QFT in curved 
spacetime will also give a rigorous result in the high energy regime.

In the BdG rediagonalization analysis, 
we assumed that the quasiparticle vacuum ($|0\rangle_{(1)}$) is time independent,
as is usual in QFT in curved spacetime.
If the quasiparticle vacuum evolves in the same manner as the condensate and,
at the final time, the initial vacuum becomes the final vacuum 
($|0\rangle_{(1)}=|0\rangle_{(2)}$), then particle creation will not occur
in the BEC system, at least at zero temperature~\cite{Yamanaka}.
In Kita theory, it is not clear how to represent the quasiparticle vacuum
~\footnote{ In Bogoliubov theory, the quasiparticle vacuum is explicitly shown in Ref. \cite{GA59,G98}.}. 
To verify this assumption, a deeper understanding of the 
quasiparticle vacuum and the Bose condensate state is required, in addition to further experimental investigation.

In this article, we used Kita theory as a conserving gapless mean field theory.
There is another theory formulated by Yukalov which also satisfies conservation laws and has gapless excitations
\cite{Yukalov1,Yukalov2,Yukalov3}.
On the basis of this theory, 
particle creation with quantum backreaction will be able to be discussed.

{\it Acknowledgments:} 
We would like to thank T.~Kita and Y.~Yamanaka for useful discussions.
This work was supported by KAKENHI, No. 21740196, No. 22740219, 
No. 21340104 and No. 19540305.

\appendix

\section{quasiparticle field expansion}
\label{sec:epsilon}

We consider a field expansion with two different compete sets
$\{{\bf u}_k^{(1)}, {\bf v}_k^{(1)}, {\bf \varphi}_0^{(1)}\}$ and 
$\{{\bf u}_k^{(2)}, {\bf v}_k^{(2)}, {\bf \varphi}_0^{(2)}\}$. 
The quasiparticle field operator can be expanded in terms of the two 
complete sets as
\begin{eqnarray}
\left(
\begin{array}{c}
\phi \\ \phi^{\dagger}
\end{array}
\right)
 &=& \sum_{k\ne0}
\left[
\hat{\gamma}_k^{(i)} \left(
\begin{array}{c}
u_{k}^{(i)} \\ -v_{k}^{(i)*} 
\end{array} \right)  + 
\hat{\gamma}_k^{(i)\dagger} \left(
\begin{array}{c}
-v_{k}^{(i)} \\ u_{k}^{(i)*} 
\end{array} \right)
\right] \nonumber \\
&& + (\hat{\gamma}_0^{(i)}+\epsilon  \hat{\gamma}_0^{(i)\dagger}) 
\left(
\begin{array}{c}
\varphi_{0}^{(i)} \\ -\varphi_{0}^{(i)*} 
\end{array} \right),
\label{eq:quaiparticle-expansion-at-each-time}
\end{eqnarray}
for $i=1,2$.
We have introduced $\epsilon = \pm1$ in the last term of the expansion equation
(\ref{eq:quaiparticle-expansion-at-each-time}), 
which is what we seek to determine in this Appendix. 

The orthonormal relations for these states are 
given by Eq. (\ref{eq:eq-ortho-bogo}).
By using these relations, we can relate $\hat{\gamma}_k^{(1)}$ and 
$\hat{\gamma}_j^{(2)}$ as
\begin{eqnarray}
\hat{\gamma}_k^{(2)} &=& \sum_{j\ne0}
\left[
\hat{\gamma}_j^{(1)} \langle {\bf u}_k^{(2)} | \hat{\tau}_3 |{\bf 
u}_j^{(1)} \rangle
+ \hat{\gamma}_j^{(1)\dagger}\langle {\bf u}_k^{(2)}|\hat{\tau}_3
| {\bf v}_j^{(1)} \rangle \right] \nonumber \\
&& + (\hat{\gamma}_0^{(1)}+\epsilon 
\hat{\gamma}_0^{(1)\dagger}  )
\langle {\bf u}_k^{(2)} |\hat{\tau}_3 | 
\mbox{\boldmath $\varphi$}^{(1)}_0  \rangle, 
\label{eq:gamma(1)-2-bogo}
\end{eqnarray}
and
\begin{eqnarray}
\hat{\gamma}_k^{(2)\dagger} &=& 
-\sum_{j\ne0}\left[
\hat{\gamma}_j^{(1)} \langle {\bf v}_k^{(2)} | \hat{\tau}_3 |{\bf u}_j^{(1)}\rangle
+ \hat{\gamma}_j^{(1)\dagger}\langle {\bf v}_k^{(2)}|\hat{\tau}_3 | 
{\bf v}_j^{(1)} \rangle \right] \nonumber \\
&&  - (\hat{\gamma}_0^{(1)}+\epsilon 
\hat{\gamma}_0^{(1)\dagger}  )
\langle {\bf v}_k^{(2)} |\hat{\tau}_3 | 
\mbox{\boldmath $\varphi$}^{(1)}_0  \rangle. 
\label{eq:gamma(1)dagger-2-bogo}
\end{eqnarray}
The bra--kets in the preceding expressions satisfy the following relations:
\begin{eqnarray}
&&\langle {\bf u}_k^{(2)}|\hat{\tau}_3
| {\bf u}_j^{(1)} \rangle =
-\langle {\bf v}_k^{(2)} | \hat{\tau}_3 |{\bf v}_j^{(1)}\rangle^*, \\
&& \langle {\bf u}_k^{(2)}|\hat{\tau}_3
| {\bf v}_j^{(1)} \rangle =
-\langle {\bf v}_k^{(2)} | \hat{\tau}_3 |{\bf u}_j^{(1)}\rangle^*, \\
&& \langle {\bf u}_k^{(2)} |\hat{\tau}_3 | 
\mbox{\boldmath $\varphi$}^{(1)}_0  \rangle
= \langle {\bf v}_k^{(2)} |\hat{\tau}_3 | 
\mbox{\boldmath $\varphi$}^{(1)}_0  \rangle^*.
\end{eqnarray}
The Hermite conjugate of Eq. (\ref{eq:gamma(1)-2-bogo}) should be
given by Eq. (\ref{eq:gamma(1)dagger-2-bogo}), 
leading to
\begin{eqnarray}
\hat{\gamma}_0^{(1)\dagger}+\epsilon \hat{\gamma}_0^{(1)} 
= -(\hat{\gamma}_0^{(1)}+\epsilon 
\hat{\gamma}_0^{(1)\dagger}).
\end{eqnarray}
Therefore, it must be 
\begin{eqnarray}
\epsilon = -1.
\end{eqnarray}

\section{Equations for mode functions}
\label{sec:equation-mode}

To derive Eq. (\ref{eq:wave-equation-metric-form}),
we write the equations for $\tilde{u}_j+ \tilde{v}^*_j$ and 
$\tilde{u}_j- \tilde{v}^*_j$.
The equation for the condensate wave function (Eq. (\ref{eq:GP-Kita})) leads to
\begin{eqnarray}
\tilde{W} 
&=& \tilde{L}+2n_0 U_0 +\hbar \partial_t S \nonumber \\
&=& K' + n_0 U_0 -i\hbar D_v -2U_0 \langle \tilde{\phi}^{\dagger} \tilde{\phi}\rangle
\nonumber \\
&& -  \frac{1}{2} U_0 \left( \langle \tilde{\phi}\tilde{\phi} \rangle 
+\langle \tilde{\phi}^{\dagger} \tilde{\phi}^{\dagger} \rangle  \right), 
\end{eqnarray}
where we have defined $K' =K - n_0^{-1/2} K n_0^{1/2}$ and 
\begin{eqnarray}
D_v : ={\bf v}_0 \cdot \nabla + \frac{1}{2}(\nabla \cdot {\bf v}_0).
\end{eqnarray}
The BdG equations (\ref{eq:BdG-mode-u-v}) then give 
equations for the function $\tilde{u}_j+ \tilde{v}^*_j$
\begin{eqnarray}
&& i\hbar (\partial_t +D_v) (\tilde{u}_j+ \tilde{v}_j^*) \nonumber \\
&\approx& 
\left(2n_0U_0  - \frac{1}{2}(\langle\tilde{\phi}\tilde{\phi}\rangle 
+ \langle \tilde{\phi}^{\dagger} \tilde{\phi}^{\dagger}\rangle  )
\right)(\tilde{u}_{j}-\tilde{v}_{j}^*) \nonumber \\
&& - U_0 \langle \tilde{\phi}^{\dagger}\tilde{\phi}^{\dagger}\rangle \tilde{u}_{j}
+ U_0 \langle \tilde{\phi}\tilde{\phi}\rangle \tilde{v}_{j}^*,
\label{eq:mode-BdG-approx-sum}
\end{eqnarray}
and for $\tilde{u}_j- \tilde{v}^*_j$
\begin{eqnarray}
&& i\hbar (\partial_t + D_v) (\tilde{u}_{j}-\tilde{v}_{j}^* )
=
K' (\tilde{u}_{j}+\tilde{v}_{j}^*)
\nonumber \\
&&\qquad - \frac{1}{2} U_0
\left( \langle \tilde{\phi} \tilde{\phi}\rangle - \langle \tilde{\phi}^{\dagger}
\tilde{\phi}^{\dagger} \rangle \right)
(\tilde{u}_{j}-\tilde{v}_{j}^*),
\label{eq:mode-BdG-diff}
\end{eqnarray}
where we have used the hydrodynamic approximation
\begin{eqnarray}
2n_0 U_0 + K' \approx 2n_0 U_0.
\end{eqnarray}

The extended GP equation (Eq. \ref{eq:GP-Kita}) 
leads to the following relations for any function $F$
\begin{eqnarray}
& (\partial_t+ D_v) \sqrt{n_0} F = \sqrt{n_0} 
(\partial_t+ {\bf v}_0\cdot \nabla) F \nonumber \\
& \qquad \qquad 
+ \frac{U_0}{2i\hbar} \sqrt{n_0}F
\left(  \langle \tilde{\phi}\tilde{\phi} \rangle - \langle \tilde{\phi}^{\dagger}
\tilde{\phi}^{\dagger} \rangle \right),
\label{lemma-1} 
\end{eqnarray}
\begin{eqnarray}
&(\partial_t +D_v) \frac{1}{\sqrt{n_0}}F
= \frac{1}{\sqrt{n_0}}
\left(
\partial_t + \nabla\cdot {\bf v}_0
\right)F  \nonumber \\
& \qquad \qquad  
-  \frac{U_0}{2i\hbar\sqrt{n_0}}
\left( \langle \tilde{\phi}\tilde{\phi}\rangle -
\langle \tilde{\phi}^{\dagger} \tilde{\phi}^{\dagger} \rangle
\right) F, 
\label{lemma-2}
\end{eqnarray}
\begin{eqnarray}
& K'(\sqrt{n_0}F) = - \frac{\hbar^2}{2m} \frac{1}{\sqrt{n_0}}
\nabla \cdot (n_0 \nabla F).
\label{lemma-3}
\end{eqnarray}

Eq. (\ref{lemma-1}) can be applied to Eq. (\ref{eq:mode-BdG-approx-sum}), 
leading to
\begin{eqnarray}
 \tilde{u}_{j}- \tilde{v}_{j}^*
& =&
\frac{i\hbar \sqrt{n_0}}{2n_0 U_0 X} 
 (\partial_t+ {\bf v}_0\cdot \nabla)
\frac{\tilde{u}_{j}+\tilde{v}_{j}^*}{\sqrt{n_0}}.
\end{eqnarray}
From this equation, we obtain Eq. (\ref{eq:f-bar-u-v}). 
Substituting this expression into Eq. (\ref{eq:mode-BdG-diff}), we obtain
\begin{eqnarray}
&& - (\partial_t+D_v)
\frac{\hbar^2}{U_0 X\sqrt{n_0}}
(\partial_t+ {\bf v}_0\cdot \nabla)
\frac{\tilde{u}_{j}+\tilde{v}_{j}^*}{2\sqrt{n_0}} \nonumber \\
& =& K' (\tilde{u}_{j}+\tilde{v}_{j}^*) 
- \frac{U_0}{2}
 \left(\langle \tilde{\phi}\tilde{\phi}\rangle -
\langle \tilde{\phi}^{\dagger} \tilde{\phi}^{\dagger} \rangle\right)
(\tilde{u}_{j} - \tilde{v}_{j}^*). \nonumber \\
\end{eqnarray}
Then, Eq. (\ref{lemma-2}) leads
\begin{eqnarray}
&& \frac{\hbar^2}{i} \frac{1}{\sqrt{n_0}}
\left(
\partial_t + \nabla\cdot {\bf v}_0
\right)\frac{m}{U_0X} 
(\partial_t+ {\bf v}_0\cdot \nabla)
\frac{\tilde{u}_{j}+\tilde{v}_{j}^*}{2mi\sqrt{n_0}} 
\nonumber \\
&=& 
K' (\tilde{u}_{j}+\tilde{v}_{j}^*).
\end{eqnarray}
The right-hand side can be rewritten using Eq. (\ref{lemma-3})
\begin{eqnarray}
K' (\tilde{u}_{j}+\tilde{v}_{j}^*) = \frac{\hbar^2}{i}
\frac{1}{\sqrt{n_0}}\nabla\cdot \left(
n_0 \nabla \frac{\tilde{u}_{j}+\tilde{v}_{j}^*}{2mi\sqrt{n_0}}
\right).
\end{eqnarray}
Therefore, the mode functions given in Eq. (\ref{def:bar-f-j})
satisfy the wave equation
\begin{eqnarray}
-(\partial_t + \nabla\cdot {\bf v}_0) \frac{m}{U_0X} 
(\partial_t+ {\bf v}_0\cdot \nabla) \bar{f}_{j}  +\nabla \cdot n_0\nabla
\bar{f}_{j} = 0, \nonumber \\
\end{eqnarray}
which can be rewritten in the form of Eq. (\ref{eq:wave-equation-metric-form})
with the effective metric (\ref{eq:effectivemetric}).


\begin{thebibliography}{99}



\bibitem{BD}
See, for example, N. D. Birrell and P. C. W. Davies, 
\lq\lq{\it Quantum Fields in Curved Space}\rq\rq,
Cambridge Univ. Press (1982).

\bibitem{hawking}
S. W. Hawking,
{\it Nature} {\bf 248}, 30 (1974);
Commun. Math. Phys. {\bf 43}, 199 (1975) [Erratum-ibid.  {\bf 46}, 206 (1976)].


\bibitem{Wald}
See, for example, R.~M.~Wald,
\lq\lq{\it Quantum Field Theory in Curved Spacetime and Black Hole Thermodynamics}\rq\rq,
The University of Chicago Press (1994).


\bibitem{Maia:2007vy}
  C.~Maia and R.~Schutzhold,
  Phys.\ Rev.\  D {\bf 76}, 101502 (2007)
  [arXiv:0706.4010 [gr-qc]].


\bibitem{Schutzhold:2008zzc}
  R.~Schutzhold and C.~Maia,
  J.\ Phys.\ A  {\bf 41}, 164065 (2008).






\bibitem{unruh81}
W. G. Unruh,
Phys. Rev. Lett. {\bf 46}, 1351 (1981).

\bibitem{novello2002}
{\it Artificial Black Holes},
edited by M. Novello, M. Visser, and G. Volovik
(World Scientific, 2002).

\bibitem{Barcelo:2005fc}
C. Barcel\'{o}, S. Liberati and M. Visser,
Living Rev. Rel.  {\bf 8}, 12 (2005).




\bibitem{Ralf05}
R. Sch\"{u}tzhold, M. Uhlmann, Y.~Xu and U.~R.~Fischer,
Phys. Rev. D {\bf 72}, 105005 (2005).


\bibitem{Balbinot:2004da}
  R.~Balbinot, S.~Fagnocchi, A.~Fabbri and G.~P.~Procopio,
  Phys.\ Rev.\ Lett.\  {\bf 94}, 161302 (2005).


\bibitem{Balbinot:2004dc}
  R.~Balbinot, S.~Fagnocchi and A.~Fabbri,
  Phys.\ Rev.\  D {\bf 71}, 064019 (2005).

\bibitem{Balbinot:2006ua}
  R.~Balbinot, A.~Fabbri, S.~Fagnocchi and R.~Parentani,
  Riv.\ Nuovo Cim.\  {\bf 28}, 1 (2005).








\bibitem{Anderson-95}
M. H. Anderson, J. R. Ensher, M. R. Matthews, C. E. Wieman, and E. A. Cornell,
Science {\bf 269}, 198 (1995).

\bibitem{Ketterle95}
K. B. Davis, M.-O. Mewes, M. R. Andrews, N. J. van Druten, D. S. Durfee, D. M. Kurn, and W. Ketterle,
Phys. Rev. Lett. {\bf 75}, 3969 (1995).



\bibitem{Garay-PRL}
L. J. Garay, J. R.~Anglin, J. I. Cirac and P. Zoller,
Phys. Rev. Lett.  {\bf 85}, 4643 (2000).


\bibitem{Lahav09}
O. Lahav, A. Itah, A. Blumkin, C. Gordon, J. Steinhauer,
arXiv:0906.1337.



\bibitem{Kurita:2007bv}
  Y.~Kurita and T.~Morinari,
  Phys.\ Rev.\  A {\bf 76}, 053603 (2007).


\bibitem{Leonhardt}
U.~Leonhardt, T.~Kiss, P.~Ohberg,
J.~Opt.~B:
Quantum Semiclassical. Opt. {\bf 5}, S42 (2003).

\bibitem{Macher:2009tw}
  J.~Macher and R.~Parentani,
  Phys.\ Rev.\  D {\bf 79}, 124008 (2009).


\bibitem{Macher:2009nz}
  J.~Macher and R.~Parentani,
Phys. Rev. A {\bf 80}, 043601 (2009).

\bibitem{Recati}
A.~Recati, N.~Pavloff and I.~Carusotto,
Phys. Rev. A {\bf 80}, 043603 (2009).


\bibitem{Braaten-Nieto1997}
E.~Braaten and A.~Nieto, 
Phys Rev B {\bf 56}, 14745 (1997). 

\bibitem{Andersen-Braaten1999}
J.~O.~Andersen and E.~Braaten, 
Phys Rev A {\bf 60}, 2330 (1999).


\bibitem{Buljan05}
H.~Buljan, M.~Segev and A.~Vardi,
Phys.~Rev.~Lett. {\bf 95}, 180401 (2005).


\bibitem{Kurita09}
Y. Kurita, M. Kobayashi, T. Morinari, M. Tsubota and H. Ishihara,
Phys. Rev. A {\bf 79}, 043616 (2009).


\bibitem{Jannes10}
C. Barcel\'{o}, L. J. Garay, and G. Jannes,
Phys. Rev. D {\bf 82}, 044042 (2010).


\bibitem{HP}
N.~M.~Hugenholtz and D.~Pines, 
Phys.~Rev.~{\bf 116}, 489 (1959).


\bibitem{Kita}
T. Kita,
J. Phys. Soc. Jpn. {\bf 74}, 1891 (2005).


\bibitem{Kita2}
T. Kita,
J. Phys. Soc. Jpn. {\bf 75}, 044603 (2006).

\bibitem{Kita3}
T. Kita,
J. Phys. Soc. Jpn. {\bf 74}, 3397 (2005).

\bibitem{Kita4}
T. Kita,
Phys.~Rev.~B {\bf 80}, 214502 (2009).




\bibitem{Yamanaka}
Y.~Yamanaka, private communication.



 \bibitem{GA59}
 M.~Girardeau and R.~Arnowitt,
 Phys.~Rev. {\bf 113}, 755 (1959).


 \bibitem{G98}
 M.~D.~Girardeau, 
 Phys.~Rev.~A {\bf 58}, 775 (1998).






\bibitem{Yukalov1}
V.~I.~Yukalov and H.~Kleinert,
Phys.~Rev.~A. {\bf 73}, 063612 (2006).


\bibitem{Yukalov2}
V.~I.~Yukalov,
Ann.~Phys. {\bf 323}, 461 (2008).


\bibitem{Yukalov3}
V.~I.~Yukalov,
Phys.~Lett.~A {\bf 359}, 712 (2006).









\end{thebibliography}
\end{document}